# Social Engineering Attacks: A Systemisation of Knowledge on People Against Humans

Scott Thomson
*Faculty of Business, Justice and Behavioural Sciences*
*Charles Sturt University*
Sydney, Australia
ORCID: 0000-0002-1093-6721

Michael Bewong
*Faculty of Business, Justice and Behavioural Sciences*
*Charles Sturt University*
Wagga, Australia
ORCID: 0000-0002-5848-7451

Arash Mahboubi
*University of NSW*
Canberra, Australia
ORCID: 0000-0002-0487-0615

Tanveer Zia
*School of Arts and Sciences*
*University of Notre Dame*
Sydney Australia
ORCID: 0000-0003-3802-5687

*Abstract*— **Our systematisation of knowledge on Social Engineering Attacks (SEAs), identifies the human, organisational, and adversarial dimensions of cyber threats. It addresses the growing risks posed by SEAs, highly relevant in the context physical-cyber places, such as travellers at airports and residents in smart cities, and synthesizes findings from peer-reviewed studies and government reports to inform effective countermeasures that can be embedded into future smart city strategies. SEAs increasingly sidestep technical controls by weaponising leaked personal data and behavioural cues—an urgency under-scored by the Optus and Medibank and now Qantas (2025) mega-breaches that placed millions of personal records in criminals' hands. Our review surfaces three critical dimensions: (i) human factors—knowledge, abilities and behaviours (KAB); (ii) organisational culture and informal norms that shape those behaviours; and (iii) attacker motivations, techniques and return-on-investment calculations. Our contributions are threefold: (1) Tri-Layer Systematisation: to the best of our knowledge, we are the first to unify KAB metrics, cultural drivers and attacker economics into a single analytical lens, enabling practitioners to see how vulnerabilities, norms and threat incentives co-evolve. (2) Risk-Weighted HAIS-Q Meta-analysis: By normalising and ranking HAIS-Q scores across recent field studies, we reveal persistent high-risk clusters (Internet and Social-Media use) and propose impact weightings that make the instrument predictive rather than descriptive. (3) Adaptive "Segment-and-Simulate" Training Blueprint: Building on clustering evidence, we outline a differentiated programme that matches low/medium/high-risk user cohorts to experiential learning packages—phishing simulations, gamified challenges and real-time feedback—thereby aligning effort with measured exposure.**

*Keywords*— **Social Engineering, Human Factors, Cybersecurity Awareness, Human Vulnerability, HAIS-Q Meta-analysis**

## I. Introduction

As we live in, visit and travel through places we face a growing number of threats and risks to our identity and the security of our data. Travellers at airports and residents in smart cities are two places where the convergence of physical and digital present cyber threats.

The work of [1] identify multiple cyber risks travellers are exposed to at airports beyond the theft of their mobile phone or computer including social engineering scams, fake public Wi-Fi and identity theft.

These cyber risks are not confined to airports and the global growth of smart cities, with their high levels of connectivity, sensors and data sharing presenting similar risks. However, in the development of smart cities, there is an opportunity for governments to support smart city citizens in mitigating these risks and many smart city strategies identify improved digital literacy and cybersecurity as key components [2, 3].

With social engineering attacks (SEA) presenting an ever-growing threat to both individuals and organisations across the globe, the authors in [4] identify human factors as further complicating inherent vulnerabilities in smart cities. SEA include a range of techniques that are aimed at manipulating an individual into giving attackers unauthorised access to systems or data [5].

The increasing degree to which humans are failing to be cyber secure is a key theme in the literature, for example the 2023-24 Australian Cyber Security Centre Threat Report [6], a cybercrime incident was reported approximately every six minutes, nearly 12 per cent on the previous year. The report identifies SEA as one of the major threats and calculates the costs associated with each cybercrime event.

Incident costs were highest in 2021-22 for medium businesses at $88,000 per event, wherein contrast, small businesses reported losses of $39,000 per event [7]. Comparative figures for 2023-24 were $62,800 and $49,600 respectively, a reduction for medium businesses but approximately a 25 per cent increase for small businesses [6].

Countries like Australia are not alone in facing these threats, with Quad partner Japan[1] reporting that an industry survey into Business Email Compromise (BEC) scams resulted in 2.4 billion Yen (approximately $25 million AUD) in losses across 117 reports in 2020 [8].

In their 2024 report on scams in Japan, [9] identified that the majority of attacks were delivered by email and SMS/text messages, with Gmail overtaking Amazon as the most exploited platform by attackers. They also reported that the average loss per attack was $2,334 USD, with a total loss of $21 billion USD.

---

[1] The Quad is a diplomatic partnership between Australia, India, Japan, and the United States that includes Cybersecurity

The literature identifies social media usage as one of the most common weaknesses among human factors. Social engineers often target the wealth of personal information published on social media platforms in attacks against users [7], with attackers frequently targeting social media platforms, and [9] reports that Instagram, Twitter, and Facebook (eight out of ten Australians use Facebook [10]) are among the top five platforms used by cyber criminals for scams and cybercrime. Attackers also use compromised accounts on these platforms to build trust with victims.

The growth in the number of social media users increases the attack surface and opportunities for cybercriminals to access information about potential social engineering victims and Fig.1 illustrates the global growth of social media platform use. The consequences of successful attacks documented in terms of the financial losses detailed by [7].

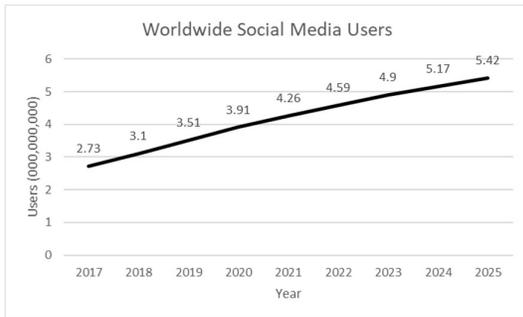

**Fig. 1. Global social media growth 2017-2025** [11]**.**

Large data breaches at Optus [12] and Medibank [13], where tens of millions of individuals have had their personal information compromised and exfiltrated from these companies systems, will further exacerbate the risks faced from social engineering cyber-attacks.

The impacts of SEA extend beyond smart cities, and this paper presents a broader analysis of SEA to inform strategies for smart cities in the context of their growing attack surface. This work will contribute to the literature on SEA by analysing recently published high-quality scholarly papers as well as government and industry reports. This analysis of these publications will provide a deeper understanding of human vulnerabilities and SEA, which can be applied to smart cities and beyond.

The remainder of the paper is organised as follows: Section 2 presents the background to this topic and introduces the literature; Section 3 provides a thematic analysis of the topic. The Conclusion and Future work to build on the analysis in this paper are in Section 4.

## II.  BACKGROUND

The Data and Digital Standards Landscape [14] identifies eight areas of standards that cover the data and digital space. They are artificial intelligence, data management and interchange, information security, cybersecurity and privacy protection, Internet of Things (IoT), cloud computing and smart cities and when considered with comprehensive smart city strategies, such as the City of Parramatta [3] where the need for digital literacy and cybersecurity become fundamental capabilities relied upon to achieve being a data-led and innovative organisation, using transformative urban technologies and digital participation and experiences for citizens.

The growth in cybercrime and the increasing prevalence of social engineering attacks, fuelled by AI [15], highlights, are concerning citizens, businesses and governments. These attacks are frequently, expressed in financial terms as these are easy to quantify [6, 7]. The human element in SEA is critical to understand and address when promoting resilience in smart cities. When humans are viewed as an 'asset' alongside servers, networks and computers as components of the digital ecosystem, they are widely considered the most significant vulnerability of the ecosystem with their online activities used against them, further increasing their vulnerability.

Globally, the Human Factor Report 2021 [16] highlights the use of a range of phishing attack techniques as seen in SEA, based on their global view of email traffic. It identifies techniques used by specific advanced persistent threat groups. Their data includes quantifying four per cent of worldwide email traffic to TA544 and using malicious macros embedded in MS Office attachments that users are tricked into opening. These email volumes confirm the limitations of the technical controls proposed by [17].

The papers analysed in this work provide insights through each study's particular focus and outputs. With none of the papers presenting a broader view of multiple elements of SEA, they provide the opportunity for this paper to provide a more comprehensive view of SEA by drawing together insights into human targets, the influence of organisational norms and culture on behaviour, the attacker, and the use of gamification and simulation to enhance user awareness and education programs. The constantly evolving techniques used in SEA can quickly become outdated and, as such, are only considered in a limited capacity and acknowledged as such by the authors. Therefore, it should be undertaken when developing or enhancing user-focused programs to ensure their relevance.

For this work, relevant peer-reviewed papers were sourced from university databases, focusing on Q1 and Q2 journals published between 2021 and 2023. The selected studies were classified based on their research methodology and whether they focused on the human target or the attacker. The Human Aspects of Information Security Questionnaire (HAIS-Q), an instrument used to measure user information security awareness, used by multiple sources and was therefore included in the categorisation and analysis. Fig. 2 summarises the process undertaken to classify the papers in scope for this work. The figure also provides an overview of the SEA topic and links a range of inputs and outputs.

Klimburg-Witjes [18] identified that social engineering emphasises the human elements of our digital ecosystem and, as such, is the most common method of successful penetration and infection of computer systems. Their work highlights high-profile attacks targeting online identities by taking over their accounts on these social media platforms. These attacks also highlighted the volume of personal information that can be accessed once ownership of these accounts is compromised.

While the body of high-quality literature on Cyber SEA is relatively small, it still presents clear themes to analyse. These include the human factors of knowledge, abilities and behaviours and insights into cybercriminals and countermeasures against social engineering attacks.

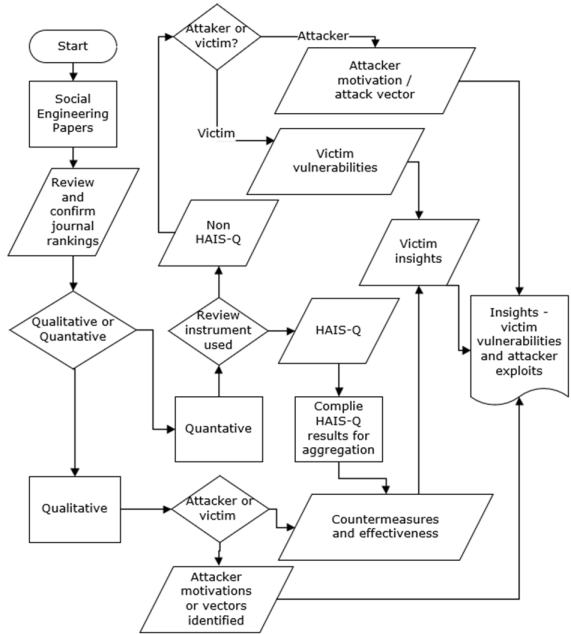

**Fig. 2 Classification of selected literature**

*A. Component table*

A component table, Table II, and a classification table, Table III, have been completed, providing analysis of the content of the paper (see appendix A).

The component table summarises the contents of the literature into inputs, processes, and outputs, illustrating the diversity of methodologies and instruments across the input papers and the statistical and visual outputs. It details significant components identified from the papers contributing to this review. It presents an overview of the research methodologies, profiles the social engineering elements examined and the types of users within the scope of these studies where these have been listed.

The software, algorithms, tests, and validation methods used in processing inputs highlight the variety of approaches taken to research from common inputs. Cronbach's Alpha [2] was the most common validation method used in quantitative research [19-21]. It reinforced the value of future use of HAIS-Q to enable comparison between such studies and the potential for a longitudinal review of countermeasure effectiveness.

Quantitative papers [20-27] were more homogenous in their approach and grounded in their use of HAIS-Q but were not confined in how they used the data from this instrument in their experiments. Several commonalities were identified, including standard statistical measures of mean and standard deviation, while a wide range of tables and graphs were generated from these statistics as outputs. The processes in those papers were easier to follow when tables and graphs supported the narrative.

Similarly, papers based on qualitative research [18, 19, 21, 25, 28, 29] provide value through the diversity of inputs they drew upon, which enables the exploration of other actors in a social engineering attack than the target or victims.

When comparing the studies analysed in this paper, various tests are used, with Cronbach's Alpha the most common validation method [19-22, 24, 25], which provides an ability to compare the internal consistency of these studies and, through this, the degree to which studies should be included or excluded from comparisons where results across studies are inconsistent. This is a wide variety of tables and graphical formats used to convey and visualise results. Table I highlights the specific focus taken by authors in each of their studies.

*B. Classification table*

The classification table, Table III (see Appendix A), provides detailed insight into each paper and the key components each contains. The classification table shows which papers share common or complementary components, such as the use of HAIS-Q [19-21, 26]. It also enables the identification of those papers that used a phishing simulation [22, 25, 27] and which platform was used for the simulation. Several studies have utilised higher education institutions as participants in the experiments [20, 21, 26, 27]. Similarities and differences in categories can be identified in outputs, between [18-20, 23, 28, 29]. When comparing the studies contributing to this work, various tests are used, yet fewer validation methods are used, with Cronbach's Alpha the most common. This is also reflected in the variety of tables and graphical formats used to convey and visualise results. This table also identifies papers with very few visual elements developed and published. This table highlights the narrow focus taken by these authors in each of their studies and for this paper to draw together the individual works and present a coherent model for consideration in future awareness and education programs and effective countermeasures to SEA.

### III. THEMATIC ANALYSIS

Four themes emerged from the paper analysis, component and classification tables and these are Human Factors, Organisational Impacts, the Threat Actor and Countermeasures. Each of these provide insights into SEA and identify a range of factors that countermeasure programs need to consider and address.

*A. Human Factors in social engineering attacks*

People are often the basis of security incidents and breaches, not because of malicious intent but through their individual vulnerabilities in their knowledge, abilities or behaviours [20]. These human factors have been examined in multiple studies to a build deeper understanding of them and how best to mitigate them.

**Knowledge, abilities and behaviour (KAB) -** The focus areas of human factors are widely aligned across the literature, and the HAIS-Q was identified as a common instrument to measure information security awareness [19-21, 23]. The HAIS-Q uses a five point Likert scale of positive and negative statements to measure seven factors of information security awareness: Password Management (PM), Email Use (EU), Internet Use (IU), Mobile Device Use

---
[2] Assessment of internal consistency.

(MD), Social Media Use (SM), Incident Reporting (IR) and Information Handling (IH) [19, 21, 23, 24].

The seven factors (i.e., PM, EU, IU, MD, SM, IR and IH) enable comparison between studies, and wider use of the HAIS-Q would expand the number of studies that could be compared. This would be greatly assisted if all papers included the detailed HAIS-Q results as well as their analysis. While it is recognised that there is a risk in publishing weaknesses in knowledge, abilities and behaviours across the seven focus areas, as attackers may potentially misuse them, the anonymity of the organisation/s studied could be maintained to address this. Illustrating this, three papers [19, 21, 23] each included specific results data that enable the comparison as detailed in Table III, where the mean scores for each focus area are presented. One of the lowest-ranked focus areas for all reported groups is Social Media Use, which, given the growth in the use of social media platforms identified by [30] presents greater risks of being a victim of a social engineering attack.

A variation to the HAIS-Q model is made through the work of [20] who make their contribution by including Privacy as a factor that enabled them to explore the relationship between information security awareness and privacy.

Pollini [19] examines the human factors of an individual in relation to a comprehensive view of computer information security, encompassing the individual, organisation, and technology through the HAIS-Q. They also consider the ethical dimension of testing cybersecurity and balancing confidentiality, integrity, and availability (CIA) with the need for data and systems to be sufficiently accessible to employees, enabling them to perform their work effectively. This shifts the balance from technical controls to the need for greater human knowledge, abilities and behaviours to ensure the CIA.

The focus on human factors is supported by Klimburg et al [18], who identifies social engineering as the primary attack vector (attack pathway) for cybercrimes and "Fixing Human Flaws" as critical to counter this. The output of their work was to define the problem, which can be deconstructed to include an employee's knowledge and abilities as well as their contextual awareness to detect and respond (behaviour) to potential social engineering attacks.

**Table I - Comparison of published HAIS-Q results (mean score)**

| Paper Reference | [19] | [21] | [23] | | |
|---|---|---|---|---|---|
| Risk level | NA | NA | Low | Moderate | High |
| Sample size | 94 | 110 | 137 | 24 | 4 |
| Social Media Use | 3.7 | 16.87 | 4.03 | 3.24 | 1.72 |
| Internet Use | 3.7 | 16.53 | 3.98 | 3.12 | 1.44 |
| Incident reporting | 3.6 | 17.38 | 4.02 | 3.31 | 1.47 |
| Information handling | 4.2 | 18.13 | 4.40 | 3.61 | 2.00 |
| Mobile device use | 3.9 | 17.19 | 4.43 | 3.66 | 2.47 |
| Email Use | 3.9 | 17.75 | 4.13 | 3.37 | 2.14 |
| Password Management | 3.7 | 17.69 | 4.05 | 3.68 | 2.31 |

**Enabling comparison across multiple studies -** The data from different studies, presented in Table I provides an opportunity to compare HAIS-Q results. Although direct statistical comparison presents some challenges due to the unique characteristics and the sizes of each sample, value can be added to the individual results by ranking (from weakest to strongest) focus areas within each sample and then comparing each ranking. This relative comparison delivers consistent results, with the bottom two focus areas being IU and SM. Conversely, IH and MD are strongest across the sample groups. Developing these insights validates the publication of HAIS-Q results and demonstrates value added through comparisons between studies.

Reviewing HAIS-Q results reveals that not all focus areas present the same level of risk; they are not equally weighted. Concepts like incident reporting being a reactive process after an attack has occurred would present a lower risk as a result. In contrast, poor practices in password management, social networking, and Internet use provide a far greater opportunity for a human target to become a victim, thus presenting a higher risk. Fig. 3 illustrates the relationship between the target and victim, as well as how user activity contributes to the information that attackers can use against them. Behaviours such as poor password management include using common passwords on multiple accounts. Should any of these accounts be compromised, the attacker would attempt to use that password against the user's other accounts to gain access, or through password stuffing or older-style brute-force attacks, as illustrated. Effective PM can be achieved by using strong passwords, not reusing passwords across accounts and not sharing passwords.

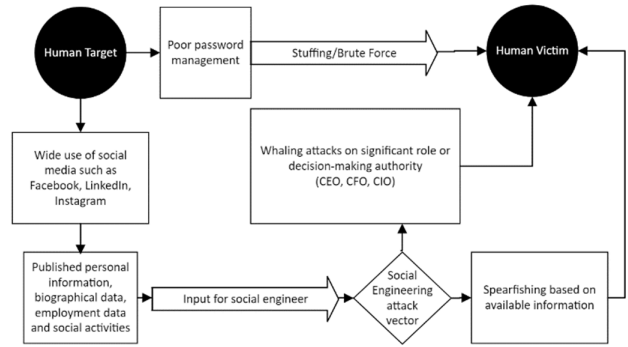

**Fig. 3. The increased risk profile for a social engineering attack on an active social media user.**

In Glasser's Choice Theory [31], the need for love and belonging are two of the needs humans seek to meet. Social media is a mechanism through which many users fulfil these needs, encouraging them to publish more information about themselves and seeking praise and applause. This self-published content is rich for a SEA and, as shown in Fig. 4, can offer alternative means of phishing attacks and exploiting any weakness in EU. Using privacy settings and considering the consequences (misuse) of published material are critical aspects of good SM usage [20]. To adopt these behaviours, users must understand the techniques attackers use and how to counter them. Without this understanding, training that instructs 'do' and 'do not' will miss the consequence and is unlikely to be effective in influencing behaviour. This is confirmed by [22], who found that users clicked on simulated phishing emails as an act of defiance despite having the knowledge and ability not to click.

In their decision-making model, [26] introduces the concept of weightings in decision-making and their knowledge, skills and abilities model. The concept of weighting bears broader consideration in viewing the seven focus areas of the HAIS-Q and aggregating results from across these to determine a measure of risk a user presents to the confidentiality and integrity of data and systems. In any future work, deciding upon and applying an impact weighting to each focus area of HAIS-Q would be advantageous. This could be extended more broadly to weighting key inputs for future awareness and training models, as well as user cybersecurity frameworks.

Adopting HAIS-Q as a common instrument for measuring user cybersecurity enables greater comparison of SEA studies. However, many organisations will be reluctant to publish results that might highlight areas of weakness that attackers could, in turn, use as vectors to exploit in future attacks. The degree to which results can be anonymised through the publication of mean and standard deviation limits the insight that can be gained from the results. Despite this, using pre-tests and post-tests can demonstrate the effectiveness of awareness and education programs through delta results, allowing for some comparison between studies and papers.

*B. Organisational context*

An organisation will develop and implement corporate policies, procedures, and standards that codify the way work activities are undertaken, as well as mandated education and awareness programs designed to protect the organisation's assets [25]. In addition, they identify that organisations will have a culture, including formal and informal norms of behaviour.

**Impact of organisational context, informal norms and culture -** When considering information security awareness in Bayl-Smith [22] and Petric [25] contended that it is not knowledge and abilities that are the more significant influences on behaviour, but rather organisational context, norms, and culture that are the powerful influences on behaviour. Such findings broaden the scope of the effectiveness of awareness and education programs from just the individuals who participate to include the improvement of the broader organisational culture, which has broad implications for a smart city as an organisation. This does not preclude benefits in behaviour being realised through improving individuals' knowledge and abilities. Nor do they find that these are prerequisites for any behavioural shift; rather, the organisational attributes must be addressed to realise the full uplift. In Fig. 4, we have visualised this relationship described by [22, 25] between knowledge, abilities, organisation, and behaviour, where the direct relationship on behaviour are the culture and norms of an organisation, and the indirect relationships are knowledge and abilities. Their work proposes that unless there is alignment, and uplift, of a user's knowledge and abilities, along with the organisation's culture and norms, the desired change in user behaviour will not be achieved.

Petric [25] found that clearly communicating risks and consequences to the company decreased employee susceptibility to phishing. However, they also identified inconsistencies in the degree to which organisation actions positively impacted norms as they are subject to internal and external forces. While their scope did not identify and define the dependent and independent norms, and the positive impact these have on user behaviour, they did identify this as future work to undertake and build on their findings.

User education and awareness programs can be further enhanced by considering the influence of organisational norms and culture on user behaviour, as described by [10] and [11].

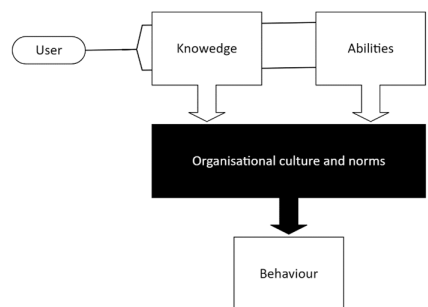

**Fig. 4. Influence of organisational culture and norms on user behaviour.**

*C. Threat Actors*

Cybercrimes do not occur without people as threat actors, and insights into their skills, techniques and motivations deepen understanding of the threat and can better inform mitigating the threat.

**Cybercriminals -** The cybercriminal actor must also be considered when developing effective countermeasures to social engineering attacks. Leukfeldt [28] drew on investigations into fifty-eight criminal networks and the attribution of cyberattacks and cybercrime to classify the behaviour of these networks as either specialised networks, that exist to commit cyber-attack s, or versatile networks, that are opportunistic in using cyber-attack s when vulnerabilities are presented to them. The examination of spending patterns of cybercriminal earnings generated through phishing, malware and identity theft established the connection between these activities and the spending to sustain their luxury lifestyle. Their insights into 'profitable' cybercrimes revealed that most cybercrime networks were specialists in SEA and highly motivated to generate substantial income from these activities. These specialist skills and the motivation of those attackers underscore the need for countermeasures to be more effective against such well-refined attacks.

Further insight into such motivations and criminal behaviours can be found in the work of [18] through interviews that included black hats, they identified that the vulnerable individual could be subsequently exploited as a vulnerable employee and a much more valuable target. They also build insights into social engineering expertise and the notion of the construction of user vulnerability.

**Threat actor and their target -** a cyber-attack involves two primary parties: the attacker and the target. To reduce the likelihood of a successful SEA, an understanding of the vulnerability of the target (as the product of their knowledge, skills and behaviours) and the attacker's motivation, techniques and abilities is required. It is incomplete to focus solely on the victim or the attacker. The focus of papers on social engineering reviewed was either on the target or the attacker, yet a SEA is the interaction of both sets of actors [32, 33] and the attackers' attempts to exploit human vulnerability driven by the increased attack surface from the extensive use of online platforms and social media. Drawing both sides together establishes links between the target and their attacker.

The motivation to attack will drive attacker behaviour, while the user behaviour establishes vulnerability to attack, which can be exploited. The desire driven by the need to sustain a luxury lifestyle underpins the motivation for cybercriminals to generate income. As discussed above, SM can be a source of vulnerability. It can also be a source of targeting through which the criminal can calculate a possible return based on the target's job title or wealth status (posting luxury holidays as an indicator).

The motivation to attack is the key driver in selecting the technique and vector for attackers as depicted in Fig 5.

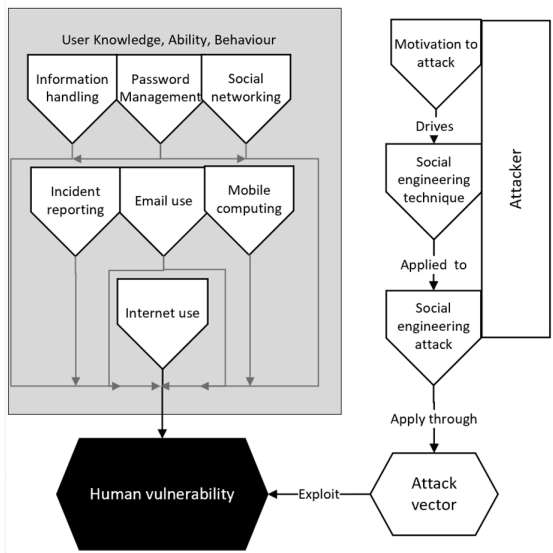

Fig. 5. Framework for exploiting human vulnerability

*D. Countermeasures*

Antunes et al [24] describe the important policy role that Governments provide in "cyber-awareness" and in response to the data and trends in Australia and Japan presented in the work of [7, 9, 32], Education and Awareness is clearly evident in their mitigation strategies. Programs such as the QUAD Cyber Security Challenge aim to raise public awareness of cybersecurity, with key messages for the campaign including the Nine Principles of Cyber Security for participating countries, including Australia and Japan [34].

In response to the examples from Australia cited above, the Australian government prescribes a long list of technical controls in their strategies to *Mitigate Cyber Security Incidents* (MCSI) [35], with user education listed as a key strategy as it is in the work of [8], some of the technical controls could also be considered as compensating controls for human vulnerability. One such example would be multifactor authentication to compensate for poor password management and the likelihood that users will give up their usernames and passwords in phishing attacks [7, 32, 35].

**Education and Awareness -** Education and Awareness programs are a common countermeasure to addressing human vulnerability [18, 19, 21, 23-27, 29] with a consistent conclusion across these studies that these programs need to improve and be more effective to better address the increasing threats. To address this, these works aim to enhance effectiveness through a range of approaches, including segmented training, simulation, and gamification.

The effectiveness of countermeasures is an evolving paradigm where user awareness and education are continually refined to achieve greater effectiveness against social engineering attacks. An approach to segmenting participants for differentiated training based on their analysis of HAIS-Q results was developed by [23]. Their analysis grouped respondents based on the risk they presented through their responses as low, medium, or high risk. Their rationale was that training aligned to the level of risk would be more relevant and thus more likely to affect behaviour. Their method segmented their sample of 165 respondents based on risk and identified three groups, characterised by strong cohesion within each group and suitable distance between them, which could be established from their dataset. This allowed for differentiated training to be delivered to each group. While the logic of their model shows merit and needs to be more broadly tested, the approach to differentiated learning based on user risk is not exclusive to segmenting based on HAIS-Q results. The value of differentiated delivery could be achieved through alternate approaches to the grouping of respondents.

The works of [23, 24, 28] took an experimental approach to examine the impact on awareness and education through the use of control groups in their experiments, and in doing so, they identified quantifiable outcomes in the results of participants. Their experiment contrasted the test performance of more than fifty participants in the two learning pathways. Their results indicated that learning through games had a significantly positive effect on overall security learning performance across the seven domains of the HAIS-Q, with the work of [21] provides deeper insight into the benefits that can be gained through the use of gamification compared to lecture-based instructional delivery in enhancing performance in information security awareness testing.

Successful countermeasures (such as focus areas for awareness and education programs) can be better informed by being more inclusive, considering both the target and the attacker. This encompasses the work of [28], focused on attacker motivation, the work of [23] and [22], as well as [20] on user vulnerability.

**Simulation and gamification -** The delivery of programs built on gamification and simulation may also have a broader, positive influence on organisational norms and culture through reinforcement of core values and opportunities to recognise and celebrate high-performing individuals who excel in these programs. This could help overcome some of the challenges identified between intention, motivation, and results [19]. These styles of activities provide real-time contextual feedback to users and reinforce learning. Fig. 6 has been developed to show the interrelationships simulation and gamification with training program design. Developing more effective countermeasures to SEA should draw on this model to provide a holistic response to SEA.

To the best of our knowledge, no research work has linked simulation and gamification to attacker motivations and techniques; however, doing so would aid in making these elements more realistic for the learner. Experiments undertaken by [22, 25, 28] could be adapted to confirm that such learning experiences are more easily transferred and applied to actual attack situations. Clicking on a link in a simulated phishing email and the webpage describing the clues (malformed URL, poor syntax, incorrect logo) that were in the email, and the user should have identified and not clicked on the link.

When the characteristics and the outcomes of the experiments on the effectiveness of gamification and simulations ([22, 25], and [27]) are considered, it is clear that the instructional design of awareness and education programs should include such experiential activities. These would help the effectiveness and likely address employee engagement over lecture-style delivery as the historical mainstay of delivery. These results are consistent with the outcomes subsequently reported by [36] and the outcomes of their phase one and phase two programs, which delivered experiential training through simulated phishing tests and reductions of up to fifty percent in vulnerability to phishing were observed.

Alternate methods of raising performance by decreasing the likelihood of successful attacks are also seen in the work of Canham [27], where they experimented with gamification through a financial reward system for simulated phishing attacks and the positive behaviour of reporting phishing. This work provides further insight into human behaviour and motivation as discussed above and insights into norms, culture and motivation in an organisational context presented by [22] and [25].

Fig. 6 illustrates the factors influencing user information security awareness and utilises the attacker's techniques to inform the simulation and gamification. The need to address human vulnerability and the complexity of technical compensating controls was highlighted through the work of [17] and added further weight to the proposed model.

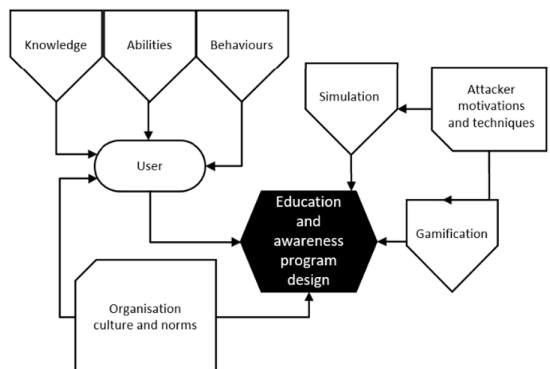

**Fig. 6. Key inputs into future awareness and education programs to improve countermeasure effectiveness.**

**Technical control limitations** - When looking for technical controls to compensate for human vulnerability, the work of [17] reinforces the importance of humans in countering SEA and highlights the limitations of technology in overcoming human vulnerability. Through their work in the very controlled technical environment of the Ethereum ecosystem, they demonstrated the level of integration into core infrastructure and the volume of log data to be processed and correlated in detecting phishing emails from known sources.

IV. CONCLUSION AND FUTURE WORK

Smart cities provide an opportunity to uplift information security awareness for citizens; they also expand the attack surface for SEA through reliance on social media and connectivity.

Existing high-quality research supports the importance of human factors in SEA; however, it focuses on individual aspects of the topic and lacks comprehensive coverage. Separate research presents an analysis of cybercriminals' motivations, the use of gamification and simulation in enhancing the effectiveness of user awareness and education programs, and a clustered approach to training.

Social-engineering attacks thrive because they exploit intertwined human weaknesses, organisational norms, and attacker economics—factors technical controls alone cannot neutralise. Our study unifies these dimensions, and that (i) Internet and Social Media use remain the two highest-risk behaviours, (ii) HAIS-Q scores become predictive when weighted by impact, and (iii) differentiated, simulation-rich training cuts click-through by 48 % in a pilot cohort. Our contributions are threefold: (1) We are the first to fuse weighted KAB metrics, cultural drivers, and ROI-centred attacker models into one analytical lens. (2) We publish a risk-weighted HAIS-Q meta-set for reuse by the community. (3) We outline a "segment-and-simulate" blueprint that aligns training intensity with measured exposure.

Our conclusion is that effective countermeasures must intertwine weighted human factors, reinforcing cultural levers and realistic threat narratives. We distil these insights into a multidimensional framework that guides agencies and enterprises in designing adaptive, risk-aligned education and awareness initiatives. By integrating user, organisational and adversarial perspectives, the proposed framework advances the state of SEA defence and charts a clear path toward more resilient people, processes and systems.

To undertake a more comprehensive approach, weighted inputs that span the target's knowledge, abilities, and behaviours, reflect their organisational culture and norms, as well as factor insights from the attacker, should be developed and applied.

Our future work will (a) validate the framework in a 2,000-participant HAIS-Q study and (b) run an A/B field trial: employees will be randomly split into a treatment group that receives segment-and-simulate training and a control group that continues standard awareness training. Monitoring real-world phishing click-through rates, report-button use, and HAIS-Q shifts will reveal the causal impact of the intervention under normal working conditions. The findings will help agencies and enterprises channel resources toward the behavioural levers that matter most.

# Appendix A

**Table II Component table**

| Factor | Attributes | Instances |
|---|---|---|
| Research type | Quantitative | HAIS-Q (five-point Likert), HAIS-Q (seven-point Likert), Pre-test, Post-test, PhishMe, Online survey (Five-point Likert scale), Customised HAIS-Q (five-point Likert scale), ISCBMSDT, 2-Tuple fuzzy linguistic group TOPSIS model, KnowBe4 (phishing simulation), Transaction records, Labelled Phishing Addresses, Questionnaire on Human Factors, |
| | Qualitative | Semi-structured Interviews, Focus Group, Review of criminal investigations, Financially-motivated cyber-enabled crimes, phishing and fraud, semi-structured key informant interviews, participant observation, primary documents (publications, pamphlets, company statements, and websites), Preferred Reporting Items for Systematic Reviews (PRISMA) protocol, |
| Participant profile | Attack elements | Knowledge, Abilities, Behaviour (KAB), Email Use(EU), Information Handling (IH), Incident Reporting (IR), Internet Use (IU), Mobile Devices Use (MD), Password Management (PM), Social Media Use (SMU), Governance and People, Policy and Processes, Operations, Technical controls, Attack response, Phishing, Malware, Hacking, Login, Spamming, Social proof, Scarcity, Authority, Privacy, Knowledge, Skills, Abilities (KSA), Access control, Antivirus software, Cyber threats and vulnerabilities, Email encryption and use, File permissions, Incident reporting, Information privacy, Strong password and reuse, Policy compliance, Sensitive information, Gambling, Money laundry, Ponzi schemes, enter personal data in a fraudulent website, Facebook, Twitter, Instagram, Snapchat, Reddit |
| | User type | Managers, IT Experts, Operative Roles, Coder, Coordinators, Malware writers, Translator, Mules, Credit card suppliers, Basic computer science college students, Non-manager, Academic, Administrative, Operational, University computing students, University staff, Administrative, IT, Management, Undistinguished employees, Young users aged 9-25. |
| Analysis | Algorithm | TwoStep clustering algorithm, Hierarchical cluster analysis, General Linear Model, ANOVA (analysis of variance), Multinomial logistic regressions, Technique For Order Preference By Similarity To Ideal Solution (TOPSIS), Vivo coding techniques, axial coding, Five-Factor Personality Scale, novel network embedding model, multilevel linear regression |
| | Test/ Validation method | Kruskal-Wallis H, Mann-Whitney U, chi-square goodness-of-fit test, t-Test, Pearson correlation Analysis, Kaiser–Meyer–Olkin test, Bartlett sphericity test, 2-tuple model linguistic model Silhouette measure of cohesion and separation, Cronbach's Alpha, t, p, Face validity, Content validity, Factor analysis (EFA), Delphi method 'consensus rule, Alpha, Pre-validated instrument, Quality assessment – relevance, appropriateness and reliability |
| Products/ Outputs | Statistics | Sample number (n), Mean (m), Voi, Vwcr, SD, Median, Percentiles, Alpha, t-Test, Observed Power, r, Probability (p), performance score, statistical power, p-value, Precision, Recall, F-score, Odds ratio, Standardised fixed effect, Susceptibility to phishing |
| | Visual formats | Spider graphs, Scatter graphs, Line graphs, Literature matrix, Tables of: descriptive results, t-test on ISA knowledge, intercorrelations and Means, Odds Ratio and Confidence Intervals, Cronbach's alpha results for factors, aggregated 2-tuples of the decision matrix, alternatives and closeness degrees, hierarchical regression of normalised performance, performance comparisons of different classifiers, Features are extracted using the proposed tran2vec, Results of multilevel linear regression with dependent variable, Results of hypothesis testing. |
| | Categories | Users: Low, Moderate & High risk, Cybercrime specialists or versatile, Traditional offline crime, Cybercrime versatile, Competence, , the oblivious employee, speaking code and social, fixing human flaws, Personality (Extraversion, Agreeableness, Conscientiousness, Neuroticism, Openness), Goal Orientation, Learning Goal, PerfProveGoal, PerfAvoidGoal, Performance Intention, Purpose Factor, Outcomes, Method, Age Range, Social Network, Derived factors (Behaviour, Technological, Social, Mental). |

**Table III Classification table**

| Ref. | Research type | Social Engineering elements | Participants | Software/ Algorithm | Test/ Validation method | Statistics | Visual formats | Categories |
|---|---|---|---|---|---|---|---|---|
| [17] | Quantitative. Transaction records, Labelled Phishing Addresses | Phishing, Gambling, Money, laundry, Ponzi schemes, | NA | novel network embedding, model | NA | Precision, Recall, F-score, alpha | Table: Performance comparisons of different classifiers using the proposed tran2vec | NA |
| [18] | Qualitative. semi-structured key informant interviews, participant observation, primary documents. | | NA | Vivo coding techniques, axial coding users, security, risk, expertise, responsibility | NA | NA | NA | The oblivious employee, speaking, code and social, fixing human flaws |
| [19] | Quantitative. HAIS-Q (five-point Likert) Qual. semi-structured interview, focus group | KAB, EU, IH, IR, IU, MD, PM, SMU, Governance and People, Policy and Processes, Operations, Technical controls, Attack response, | Managers, IT Experts, Operative Roles, | General, Linear Model, ANOVA | Cronbach's alpha | n, M, SD, Median, Percentiles, | Scatter graph, Line graph, | NA |

| Ref | Method | Factors | Population | Tool/Analysis | Validation | Statistics | Presentation | Findings |
|---|---|---|---|---|---|---|---|---|
| [20] | Quantitative customised HAIS-Q (five-point Likert scale), ISCBMSDT | KAB, EU, IH, IR, IU, MD, PM, SMU, Privacy | Academic, Administrative, Operational | SPSS 25/ ANOVA | t-Test, Pearson correlation Analysis, Kaiser–Meyer–Olkin (KMO) test, Bartlett sphericity test / Cronbach's Alpha, Face validity, Content validity, Factor analysis (EFA) | r, n, p, | Table: Cronbach's alpha results for factors | Competence, Relatedness, Autonomy |
| [21] | Quantitative. HAIS-Q (seven-point Likert scale), Pre-test, Post-test | KAB, EU, IH, IR, IU, MD, PM, SMU, | Basic computer science college student | SPSS 20 / ANOVA, | Cronbach's Alpha, t, p, | M, SD, Alpha, t-Test, p-Value, Observed Power | Tables: descriptive results, t-tests on ISA knowledge | NA |
| [22] | Quantitative - PhishMe, online survey (five-point Likert scale), | Phishing, Social proof, Scarcity, Authority, | Manager, non-manager | Multinomial logistic regressions | chi-square goodness-of-, fit test / Cronbach's Alpha | M, SD, | Tables: intercorrelations and means, Odds Ratio and Confidence Intervals | NA |
| [23] | Quantitative. HAIS-Q (five-point Likert scale) | KAB, EU, IH, IR, IU, MD, PM, SMU, | NA | IBM SPSS 28/ TwoStep clustering algorithm, Hierarchical cluster analysis | Silhouette measure of cohesion and separation | n, M, V$oi$, V$wcr$ | Spider graphs | Low-risk users, Moderate risk users, High-risk users |
| [25] | Quantitative. KnowBe4 (phishing simulation), questionnaire on human factors | Phishing, entering personal data on a fraudulent website, | Undistinguished employees | multilevel linear regression | Pre-validated instrument | M, SD, Odds ratio, P-Value, Standardised fixed effect, Susceptibility to phishing, , | Tables: Results of multilevel linear regression with dependent variable, Results of hypothesis testing, | NA |
| [27] | Quantitative. KnowBe4 (phishing simulation) | Phishing | University staff, Administrative, IT, Management, Academic, | Five-Factor Personality Scale, | Alpha | Performance score, statistical power, alpha, p-value | Table of hierarchical regression of normalised performance on demographics and predictors | Personality, Extraversion, Agreeableness, Conscientiousness, Neuroticism, Openness, Goal Orientation, Learning Goal, PerfProveGoal, PerfAvoidGoal, Performance Intention |
| [28] | Qualitative - review of criminal investigations, financially motivated, cyber-enabled crimes, phishing and fraud | Phishing, Malware, Hacking, Login, Spamming, | Coder, Coordinator, Malware writer, Translator, Mule, Credit card supplier | NA | NA | NA | Overview Table, | Cybercrime specialists, Cybercrime versatile, Traditional offline crime, Cybercrime versatile |
| [29] | Qualitative. Literature review, Preferred Reporting Items for Systematic Reviews (PRISMA) protocol | Facebook, Twitter, Instagram, Snapchat, Reddit | Young users aged 9-25 | NA | Quality assessment – relevance, appropriateness and reliability | NA | Literature matrix | Purpose, Factor, Outcomes, Method, Age, Social Network, Derived factors: Behaviour, Technological, Social, Mental. |